\documentclass{sig-alternate}

\begin{document}

\pdfoutput=1


%
\conferenceinfo{WOODSTOCK}{'97 El Paso, Texas USA}

\title{Achieving Architectural Sustainability via Diversification}
%
%
%
%
%

\title{Diversifying Software Architecture for Sustainability: \\
A Value-based Perspective}

\numberofauthors{2} 
%
\author{
%
%
\alignauthor
Dalia Sobhy\\
       \affaddr{School of Computer Science}\\
       \affaddr{University of Birmingham}\\
       \affaddr{Birmingham B15 2TT, United Kingdom}\\
       \email{dms446@cs.bham.ac.uk}
\alignauthor
Rami Bahsoon\\
        \affaddr{School of Computer Science}\\
       \affaddr{University of Birmingham}\\
       \affaddr{Birmingham B15 2TT, United Kingdom}\\
       \email{r.bahsoon@cs.bham.ac.uk}
}       

\maketitle
\begin{abstract}

Although the concept of software diversity has been thoroughly adopted by software architects for many years, yet the advent of using diversity to achieve sustainability is overlooked. We argue that option thinking is an effective decision making tool to evaluate the trade-offs between architectural strategies and their long-term values under uncertainty. Our method extends cost-benefit analysis method \textit{CBAM}. Unlike CBAM, our focus is on valuing the options which diversification can embed in the architecture and their corresponding value using \textit{real options pricing theory}. The intuitive assumption is that the value of these options can provide the architect with insights on the long-term performance of these decisions in relation to some scenarios of interest and use them as the basis for reasoning about sustainability. The method aims to answer the following: \textit{(1) Is diversification of architectural decisions beneficial in sustaining the software}, \textit{(2) When, where and to what extent.} The proposed model is illustrated and evaluated using a case study from the literature referred to as GridStix.
\end{abstract}

\section{Introduction}
Diversity, in all of its manifestations, is an essential conditioning factor for the successful evolution of various systems, summarized in the ancient dictum, `variety is the spice of life'. Providing a wide range of species forming disparate populations is generally related to the term \textit{diversity}, which includes biological, ecological, chemical, cultural, workplace diversity etc. These types of diversity aim at maximizing the benefits from varieties. For example, as in workplace diversity, having diverse groups of employees from different backgrounds would lead to greater creativity and innovation. In its simplest form, biological diversity positively impacts the quality of human existence, hence protecting the environment from extinctions \cite{huston1994biological}.    

In the past years, seemingly endless streams of academic literature, which tout the benefits of diversity in design, have filled bookshelves and the airwaves. Particularly, in 1984, the natural forms of diversity inspired J. Kelly to introduce design diversity; \textit{``the approach in which the hardware and software elements that are to be used for multiple computations are not copies, but are independently designed to meet a system's requirements"} \cite{Avizienis:1984:FTD:1319725.1320045}. In other words, it is the generation of functionally equivalent versions of a software system, but implemented differently \cite{1701972}. Further, diversity could be interpreted as a means to defend a common system from uncertainty, which rises linearly with it. In this context, software diversity raised the potential awareness to apply diversification in the decision-making process, which in turn may cause a noticeable improvement in the way we design dependable, evolvable, and sustainable software. 

Sustainability is one of the major concerns for the longevity of any system. To clarify, `longevity' as a term means how long a system will continue to adapt with the advancement of technologies and changes in user requirements and the environment. It is classically defined as \textit{``meeting the needs of the present without compromising the ability of future generations to satisfy their own needs"} \cite{brundtland1987report}. Due to the importance of sustainability attribute for the evolution and longevity of software systems, the IEEE Software for example published a special issue on architecture sustainability, which thoroughly discussed the long-term need for sustainability \cite{avgeriou2013architecture}, possible metrics for sustainability\cite{koziolek2013measuring}, and the necessary milestones to achieve it \cite{savolainen2013long}. Our main focus will be on Architecture sustainability, which generally refers to long-term cost-effective adaptation and immense emergence of a system towards diverse kinds of change, such as requirements, environment, business strategies and goals, technology, accidental complexity, and false decisions\cite{avgeriou2013architecture}.

There is a general belief that there exist a correlation between diversity and sustainability. In light of this, natural forms of diversity are prominent proof of employing more diverse system is likely to sustain more in the future. On the opposite extreme, the link between diversity and sustainability is still unexplored in software engineering. So the question now is \textit{how that link can help in engineering more sustainable software?} We argue that this link can be judged from a value-based perspective. More specifically, the focus is on how we can value diversity in architecture design decisions and evaluate their contribution to achieve sustainability in software systems. As the valuation shall take into consideration uncertainty, we appeal to options thinking to answer the above question.



In a nutshell, the novel contribution is an architecture-centric method, which builds on cost-benefit analysis method (CBAM)\cite{AsundiUsingEconomic2001} to evaluate and reason about how architectural diversification decisions can be employed and their augmentation to value creation. The approach uses real options analysis \cite{trigeorgis1996real} to quantify the long-term contribution of these decisions to value and determine how that value can assist decision makers and software architects to visualizing about sustainability in software. Scenarios from case study are necessary to demonstrate the validity and applicability of approach. Our exploratory case analysis is based on provisional data gathered from GridStix prototype, deployed at River Ribble in the North West England\cite{riiver_dataset}\cite{grace2008experiences}. 


The remainder of this paper is structured as follows: Section 2 elaborates on the motivation for proposing the approach, section 3 explains briefly the techniques used to support our method. Section 4 presents the method. Section 5 applies the approach to a case study and provides a constructive evaluation for the method. Section 6 presents the related work. Section 7 concludes the work.

\section{Motivation}


Adhering to principles of diversity, it is a key enabler for sustainability. Beyond this, however, the link between diversity and sustainability was overlooked in the context of software engineering. In particular, this link has not been studied from value-based perspective. This need can be true for many artefacts that stem from engineering. We contend that a sustainable software shall create and add value, while in operation and as it evolves. Moreover, the cost-benefit trade-offs  and long-term value have significant impact on the extent to which an architecture can evolve and sustain during its lifetime. A prominent demonstration of the issue is the Concorde, a prestigious supersonic jet that flies with up to double the speed of sound at an expensive cost. During the operation of Concorde, only two major incidents happened. A popular accident named ``Paris Crash" occurred after almost 35 years of operation where the engines flamed, resulting in more than hundred deaths. Despite its high safety record over the operation lifetime, the operation expense and its inability to sustain value were among the major reasons for phasing out Concorde.



\section{Background}
This section sets the concepts needed by the method
to quantify the impact of applying diversified decisions on system's quality attributes (QA).


\subsection{CBAM}
\begin{figure}[!t]
\centering
\includegraphics[scale=0.3]{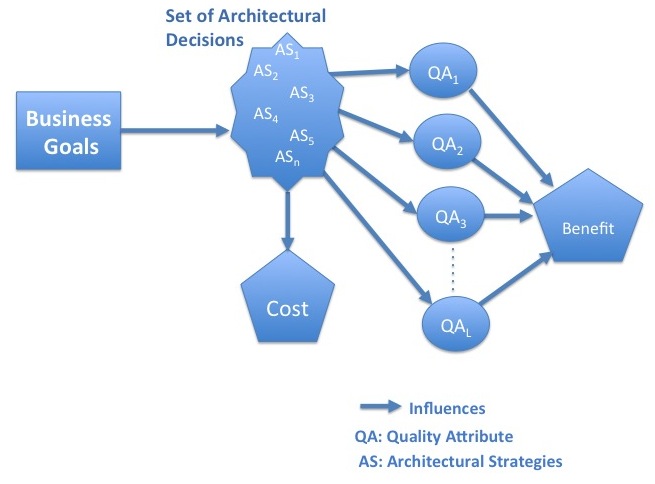}
\caption{Steps of Classical CBAM \cite{AsundiUsingEconomic2001}}
\label{fig:cbam_classic}
\end{figure}

CBAM extends ATAM\cite{kazman2000atam} with explicit focus on the costs and benefits of the architecture decisions in meeting scenarios related to QA. Furthermore, it intends to \textit{``develop a process that helps a designer choose amongst architectural options, during both initial design and its subsequent periods of upgrade, while being constrained to finite resources"} \cite{AsundiUsingEconomic2001}. The classical CBAM is illustrated in figure~\ref{fig:cbam_classic}, which starts with the determination of business goals, the fundamental for constructing a software architecture. Commonly, it has been troublesome for stakeholders to elicit mission goals, due to their presence in various levels of abstraction. A deep review for the categorization of business goals is found in \cite{kazman2005categorizing}. For example, a typical goal could be cutting down the maintenance and development costs or providing a highly reliable and secure system. In this context, these goals drive the architecture decisions (sometimes called architectural strategies AS), which have two major ramifications: technical and economical. The former implications generate the QAs, such as performance, usability, security, etc., portraying the system. While the latter are the costs of implementation and benefits obtained from the system. Finally, the ASs generate costs and QA responses, which influence the benefits \cite{AsundiUsingEconomic2001}\cite{kazman2001quantifying}.

\subsection{Real Options Analysis}
We view architecting for sustainability through diversification as an option problem. Real options theory provides an analysis paradigm that emphasizes the value-generating power of flexibility under uncertainty. An option is the right, but not the obligation, to make an investment decision in accordance to given circumstances for a particular duration into the future, ending with an expiration date \cite{trigeorgis1996real}. Real options are typically used for real assets (non-financial), such as a property or a new product design. Furthermore, it can be categorized into call and put options. The former gives the right to buy an uncertain future valued asset for the strike price by a specified date, whereas the latter offers the option to sell that asset. There are several types of options: switch, growth, abandon, defer, build, and alter.  As an example of switching to improve the system QA, if we are designing for sustaining value at runtime, the architect may think about how switching between various diversified decisions can support value creation (i.e. the option can be offering a suitable switching type). As a further illustration, a diversified option aiming to scale the system can be treated as a growth option.

Real options analysis is well-known type of option for the use in strategic decision-making scenes \cite{bowman2001real}. As a recall, we contend that diversification is likely to create decisions in the form of real options that can be exercised for value creation. In fact, the value of these options is long-term and can cross-cut many dimensions, including economical and technical ones. Particularly, the valuation of the options can be performed in accordance to the sustainability dimensions, which relate to technical, individual, economics, environment, and social \cite{becker2014karlskrona}. For this work, we focus on the technical issues that allied to QA responses and their value contribution to sustainability. 

\section{Diversified CBAM for Evaluating Architecture Sustainability with Real Options}

\begin{figure}[!t]
\centering
\includegraphics[scale=0.3]{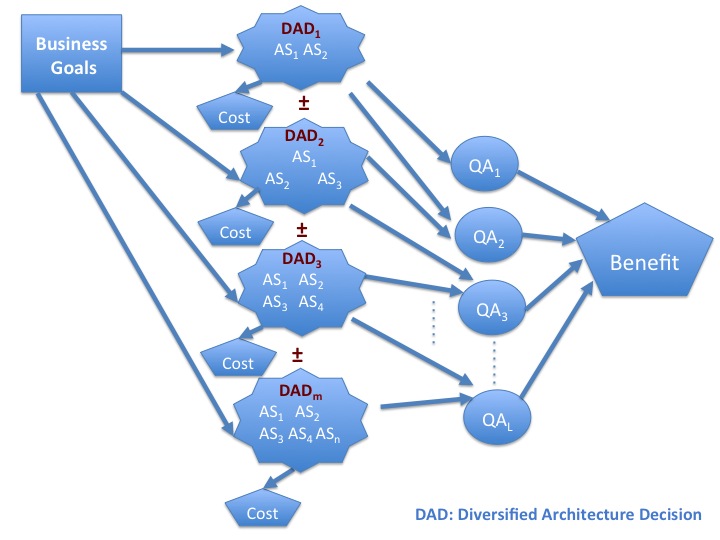}
\caption{Diversified CBAM}
\label{fig:cbam_div}
\end{figure}

The proposed approach extends CBAM with an emphasis on diversification architectural decisions (DAD), their cost and the value they add to the software, as illustrated in figure~\ref{fig:cbam_div}. More specifically, we use this information as a way to reason and reflect about sustainability. CBAM tends to quantify the extent to which an AS meets the scenarios and its response level, whereas our method operates on the improvement (if any) on the QA responses once we consider diversification. This improvement's response can carry an added value, but with a price. The use of real option can adequately help to quantify and visualize that response under uncertainty. The essence is in the future opportunities flexibility creates and its contribution to long-term value creation.Therefore, the key issue is how to select the best-fit diversified strategy according to the current goal. Our holistic view is reaching a set of decisions that can have a global impact on value and hence sustaining the system. In this section, the steps of the method are introduced.

\subsection{Procedures of Approach}


There are two fundamental questions to be answered in our method: which diversity solution would provide more favorable option and when to exercise the offered option. Despite the prominent outcome from having variety of options, sometimes the value of an option in terms of complexity exceeds its benefit. In this context, real-options theory is employed along with CBAM, for example to provide design support for run-time decisions that promote diversification, which in turn sustain the software system.

\subsubsection{Step 1: Choosing the business goals, Scenarios and DADs}

The preliminary analysis aims at providing the system with typical business drivers, a number of scenarios of interest, and varying architectural decisions, which uses the formulation of CBAM.  However, our method focuses on QA and their responses with respect to scenarios of interest, which are key tenets for sustaining the system. Moreover, in the diversified CBAM, DADs are envisaged as portfolio of options. Exercising each decision can be formulated as call option, with an exercise price and uncertain value. A collection of alternate ASs are gathered together forming a DAD, where various DADs could be implemented in the system at the same time. 

\subsubsection{Step 2: Assessing the relative importance of QAs (Elicit QAScore\textsubscript{j})}

The question that steers this step:  which QAs are of critical importance for sustaining the system and what is the relative importance? The evaluation has explicit focus on long-term scenarios of exploratory and expansion nature. The assumption is to further ``tune" the architecture so it can continue to deliver value if these materialize. Specifically, this is translated in terms of a ranking score to each QA in accordance to the system's business goals, given that we have\textit{ j} QAs, the total scores should satisfy equation (\ref{eq_QAScore}).

 
\begin{equation}
\sum \nolimits_{j}QAScore_{j} =100   ,  \forall j: QAScore_{j} \geqslant 0
\label{eq_QAScore}
\end{equation}



\subsubsection{Step 3: Quantifying the benefits of the DADs (Elicit ContribScore\textsubscript{j})}

Classically, any DAD has an intuitive effect on multitude of QAs. This typical influence results in favouring certain QAs and hurting others. An apparent constraint arise from selecting a DAD is the uncertainty about its rational impact on the system. In this context, the elicitation of DAD's effect on each QA is a critical exercise, which is denoted by \textit{ContribScore}. This score is theoretically rated using -1 to 1 scale, where `1' resembles the best impact on QA and `-1' means the opposite.

The values of QAScore and ContribScore are given as input to the method.
Although outside the scope of the method, one possibility to
derive the value \textit{ContribScore} is to employ the
Kendall's Concordance Coefficient \cite{AsundiUsingEconomic2001}. It is a notable aid in addressing the variability of scores by valuing the consistency between raters of elicited contribution scores. Low coefficient signifies high variability quota, which denotes that some stakeholders are lacking particular knowledge. 
 
The subsequent step is quantifying the benefit of each DAD \textit{(Benefit\textsubscript{DAD\textsubscript{i}})} using equation (\ref{eq:benefit_dad}).
\begin{equation}
Benefit_{DAD_{i}} =  \sum \nolimits_{j} QAScore_{j} * ContribScore_{i,j}
\label{eq:benefit_dad}
\end{equation}

\subsubsection{Step 4: Quantifying the costs of DADs and Incorporating Scheduling implications}

Classical CBAM uses the common measures for determining the costs, which involves the implementation costs only. Unlike CBAM, our approach embraces the switching costs between decisions, which is equivalent to primary payment required for purchasing a stock option. This is in addition to the costs of deploying DADs, configuration costs, and maintenance costs, similarly to the exercise price, denoted by \textit{Cost (DAD\textsubscript{i})}.  
To this extent, the cost metric is a paramount indicator for quantifying the option price, in addition to the benefit ramifications. For instance, decision A could provide high utility and high cost, and hence high risk. While decision B supplies low utility and low cost, but low risk. In this context, high costs are likely to generate a common threat towards achieving sustainability. Most importantly, the costs of all DADs deployment should not exceed the total budget of the project \textit{E} as shown in equation (\ref{eq:cost_dad}). As in CBAM, we measured the costs with respect to 1-100 range, where `100' denotes the most expensive cost and `1' is the cheapest. 
   


\begin{equation}
\sum \nolimits_{i} Cost (DAD_{i}) \leqslant E
\label{eq:cost_dad}
\end{equation}

\subsubsection{Step 5: Calculate the Return of each DAD for the scenarios}
As aforementioned, diversification decisions can be formulated as options that can be evaluated in the presence of uncertainty. Our case typically focuses on options to switch, wait and/or abandon. DADs supporting diversification can be valued as calls. The value of these calls provide the right without symmetric obligations to \textit{switch} among DADs if the exercise is favourable. Moreover, the option to \textit{wait} to switch is advantageous, the rational investigation of an option and waiting for provocative outcome has a positive impact on switching options. Last, but not least, if a DAD is no longer favouring the system, then preferably \textit{abandon(e.g. phase out, rethink its design, etc)} that DAD.

\begin{figure}[!t]
\centering
\includegraphics[scale=0.25]{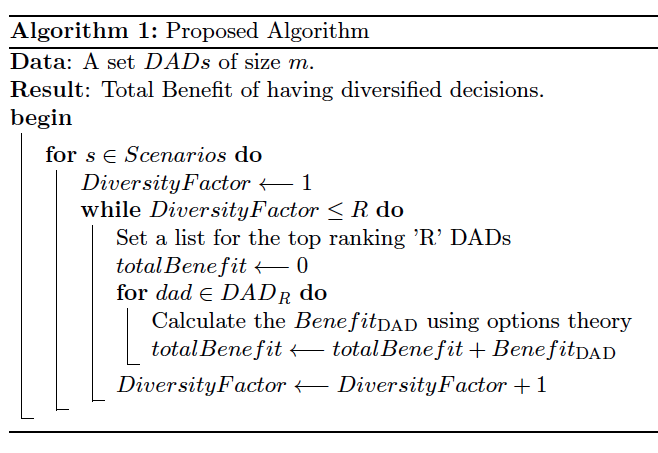}
\caption{Proposed Algorithm}
\label{fig:PR}
\end{figure}

We view DADs as portfolios with options, i.e. each DAD is a portfolio with its varying exercise cost, risks, and benefits observed on QA responses. Each DAD comprises a collection of ASs, so that similar or different ASs could be implemented in multiple DADs, as previously shown in figure~\ref{fig:cbam_div}. Moreover, figure \ref{fig:PR} illustrates the steps of applying diversification using portfolio of options where real options analysis is employed for value quantifications.


We have exploited the binomial pricing model \cite{brandao2005using} for options valuation, as well as Ozkaya et. al, approach \cite{ozkaya2007quality} for guidance. The binomial model provides a visual representation for the options. It is a constructive aid aiming to show the suitable time slot for exercising an option. For each step of binomial tree, the up and down nodes values are important in determining the system value rise and fall, which are ultimately used to calculate the option price. In our method, the impact of applying each DAD on the system is computed at every time slot \textit{t}, where t=\textit{k} indicates that the time equals to \textit{k} unit time of interest. The following steps are necessary for valuation of options using binomial option pricing model.


\begin{enumerate}
\item \textit{\textbf{Calculate the system value without factoring diversification into the decisions}}

As a start, the system value \textit{S\textsubscript{DAD}} is evaluated with respect to the initial system value denoted by \textit{V\textsubscript{s}} and resultant benefit of deploying DADs as in equation~\ref{eq:s_dad}. To exemplify, consider possessing multiple DADs, then the system value is the summation of implemented DADs  and \textit{V\textsubscript{s}}. As in equation~\ref{eq:s_dad_t}, let \textit{S\textsubscript{DAD}(t)} be the system value after implementing a particular DAD causing either incremental improvement or degradation at time \textit{t}, which is equivalent to the uncertain stock price when modelling an American call option. 

\begin{equation}
S_{DAD} = V_{s} + \sum \limits_{i}Benefit_{DAD_{i}}
\label{eq:s_dad}
\end{equation}

\begin{equation}
S_{DAD}(t) = V_{s} + \sum \limits_{i}Benefit_{DAD_{i}}(t)
\label{eq:s_dad_t}
\end{equation}

 \textit{\textbf{\item Calculate the likely rise and fall of payoff with DADs}}

 Let \textit{ f\textsubscript{u}} be the likely rise of payoff i.e. rising option price from implementing DAD, whereas \textit{f\textsubscript{d}} is the likely fall of payoff i.e.. falling option price from implementing DAD. This is typically interpreted using equations (\ref{eq_fu},\ref{eq_fd}). Besides, \textit{Cost (DAD\textsubscript{i})} is the general costs for implementing DADs as illustrated in equation~\ref{eq:cost_dad}, which could constitute the implementation of one or more DADs. Furthermore, the system value (corresponding to stock) benefiting from DAD i.e. value rise, is denoted by the up factor \textit{u}, whereas \textit{d} is the down factor denoting the system value being hurt from DAD i.e. value fall. The \textit{u} and \textit{d} coefficients are formulated with respect to the volatility of the anticipated outcome. Certainly, the formulas \textit{ f\textsubscript{u}} and \textit{f\textsubscript{d}} will never generate a negative value and elicit the maximum possible value from the net payoff, which makes them optimal. 

\begin{equation}
f_{u} = max(0,uS_{DAD}(t) - \sum \nolimits_{i} Cost (DAD_{i}))
\label{eq_fu}
\end{equation}
\begin{equation}
f_{d} = max(0,dS_{DAD}(t) - \sum \nolimits_{i} Cost (DAD_{i}))
\label{eq_fd}
\end{equation}

 \textit{\textbf{\item Calculate the option price of exercising DADs}}

A further critical step is determination of the option price of exercising diversified decisions \textit{f}, which reveals at what time \textit{t} it is favourable to take the decision i.e. exercise an option. Moreover, it also illustrates the long-term performance of a system, which in turn aids in promoting sustainability. An important coefficient \textit{p} carries the risk-adjusted probability. It is essential to note that \textit{u}, \textit{d}, and risk-free interest rate \textit{r} values are elicited from stakeholders, given the following constraint: d < 1+r < u. 


\begin{equation}
f = \frac{pf_{u} + (1-p)f_{d}} {1-r}
\end{equation}

\begin{equation}
p = \frac{1+r-d} {u-d}
\end{equation}

\end{enumerate}

To move beyond the general approach, it is necessary to perform an exploratory analysis to evaluate the proposed method via a case study, which is illustrated in the next section.

\section{Evaluation \& Discussion}
To test the applicability of approach, this section presents an application of the method on GridStix prototype\cite{hughes2006gridstix}. The evaluation is performed using hypothetical data, aiming to measure the long-term effectiveness of implementing various diversified decisions i.e. portfolio of options, on system QAs.


\subsection{GridStix}
A Grid-based technique to support flood prediction process by implementing embedded sensors with diverse networking technologies \cite{hughes2006gridstix}. Particularly, it measures the water level for flood anticipation. We have chosen GridStix, because of its great variability in design space offering a sizzling environment for the development of diversification. It is portrayed by numerous QAs and different architectural components \cite{grace2008experiences}. To this extent, the elements of design space are adequate for having diversified design decisions, which is necessary for our case evaluation. Typically, our method is aiming to envisage diversification by providing a portfolio of options, which are evaluated from value-based perspective under uncertainty, to promote sustainability. In other words, it is how to engineer the concept of diversity in software architecture. Furthermore, GridStix is one of the well-documented case studies that we found in research, illustrating the pros and cons, and hence a good aid for assessing our approach.  
	
\subsection{Application}

\subsubsection{Step 1: Choosing the business goals, Scenarios and DADs}
Among the business goals, which we consider to illustrate our approach are the accuracy of flood anticipation and reasonable warning time prior the flood. We then elicit different scenarios  -see Table \ref{table:QAresponses}- with respect to QAs and the specific concerns. Moreover, the general capabilities offered by GridStix along with the primitives for diversification is shown in Table \ref{table:capabilities}. 

\begin{table*}
\centering
\caption{QA Responses}
\begin{tabular}{|c|c|c|p{3in}|} \hline
& QA &Concern&Scenarios\\ \hline
1 & Performance&GridStix Node Latency&Messages transmission between any given sensor node and gateway should be $\leq$ 30ms\\ \hline
2 &Availability & Hardware Failure & Gateway failure detected and recovered in <1 minute \\ \hline
3 &Reliability & Flood Prediction Accuracy & The system should send alert messages in < 2 seconds for flood anticipation\\ \hline
& & Network Resilience & The average number of routes from a given sensor node to gateway > 13 \\ \hline
4 & Scalability & Data Management & if a node has full capacity, it should forward the incoming message to a neighbor node in $\leq$ 100ms (75ms for choosing the best available neighbor node and 25ms for transmission) \\ \hline
5 & Energy Efficiency& Minimize power consumption & Average power consumption for 1KB data forwarding from node to gateway $\leq$1400mW\\ \hline
6 & Security & Node Manipulation & Gateway should be 99.99\% secured against any data manipulations\\ \hline
\end{tabular}
\label{table:QAresponses}
\end{table*}

\begin{table}
\centering
\caption{General capabilities provided by the GridStix}
\begin{tabular}{|c|l|l|} \hline
& General Capabilities & Primitives for Diversification (ASs)\\ \hline

1 & Connectivity & Wifi, BT, and GPRS\\ \hline
2&Data Management & Local Storage, Shared DB, Cloud, \\
 & & Neighbour Node \\ \hline

3&Power Consumption & Built-in Battery (Rechargeable),\\
& & Battery Replacement,\\ 
& & and Energy Conservation \\
 & &via Solar Cell, Kinetic Energy\\ \hline
4& Routing Algorithm & FH, SP\\ \hline
\end{tabular}
\label{table:capabilities}
\end{table}

\subsubsection{Step 2: Assessing the relative importance of QAs (Elicit QAScore\textsubscript{j})}

The relative importance of QAs for GridStix are elicited from stakeholders, with respect to equation \ref{eq_QAScore} as shown below:

\begin{itemize}
\item Performance: 20
\item Reliability: 30
\item Availability: 20
\item Security: 10
\item Scalability: 5
\item Energy Efficiency: 15
\end{itemize}

\subsubsection{Step 3: Quantifying the benefits of the DADs (Elicit ContribScore\textsubscript{j})}

Table \ref{table:QAContribScore} is employed to determine the impact of each DAD on the QAs (\textit{ContribScore}). As a recall, this score is rated using -1 to 1 scale, where '1' resembles the best impact on QA and '-1' means the opposite. For instance, for DAD\textsubscript{1} deployment, a QA response of 1.0 reliability reveals that despite the system is highly reliable, it has poor energy efficiency of -0.2 and so on. In other words, the positive QA response should not overlook the negative impact on other QAs i.e. tradeoffs between QAs.

\begin{table*}
\centering
\caption{The impact of a DAD on QAs and DADs costs}
\begin{tabular}{|p{1in}|c|c|c|c|c|c|c|} \hline
DAD\textsubscript{i} &Performance&Reliability&Availability & Security&Scalability & Energy Efficiency & Cost\\ \hline
1- Wifi & 0.6&1.0&0.7&0.3&0.7&-0.2&30\\ \hline
2- BT & 0.1&0.4&0.9&0.8&-0.4&0.8&20 \\ \hline
3- FH & 0.5 &0.8 &0.8&0.0&0.0&-0.4&15\\ \hline
4- SH &0.2 &-0.1&0.5&0.0&0.0&0.7&10\\ \hline
5- Wifi+FH & 1.0 &1.0 &0.9 & -0.1&0.7&-0.6&45\\ \hline
6- Wifi+SP & 0.7 & 0.5 &0.5&-0.1&0.7&0.2&40\\ \hline
7- BT+FH &0.5 &0.2 &0.6&0.8&-0.4&0.7&35\\ \hline
8- BT+SP&0.2 &0.1 &-0.2&0.8&-0.4&1.0&30\\ \hline
\end{tabular}
\label{table:QAContribScore}
\end{table*}

\subsubsection{Step 4: Quantifying the costs of DADs and Incorporating Scheduling implications}

The general costs involved for each DAD are depicted in table~\ref{table:QAContribScore}, which is similar to the exercise price of an option. For better illustration, we multiplied all the costs and benefits by a factor of 25. 




\subsubsection{Step 5: Calculate the Return of each DAD for the scenarios}

Generally, our method is a scenario-based mechanism, which utilizes scenarios as a way to determine the trade-offs between QAs. Drawing on case study, the scenario employed for evaluation is: (Sc) \textit{Messages transmission between any given sensor node and gateway should be $\leq$ 30ms (addressing the performance QA)}, hence improving the average network latency provided in \cite{grace2008experiences} roughly by 60\%. The stakeholders chose the top ranked DADs, which are suitable for implementation aiming to achieve the prior objective: DAD\textsubscript{1},DAD\textsubscript{3}, DAD\textsubscript{5}, and DAD\textsubscript{7}. 

\begin{figure}[!t]
\centering
\includegraphics[scale=0.4]{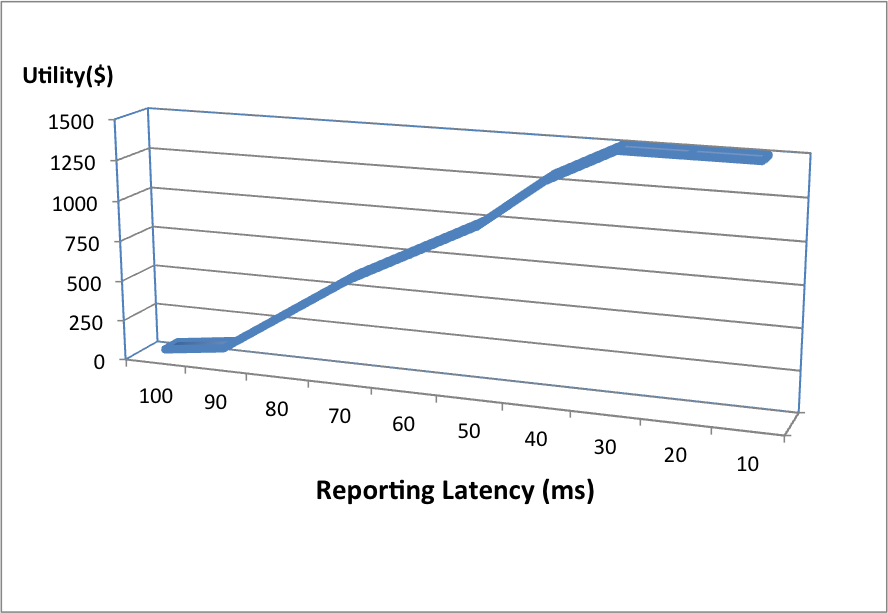}
\caption{Utility versus Reporting Latency in case of implementing additional nodes}
\label {fig:utility_latency}
\end{figure}

Currently, the approximate number of deployed gridstix nodes is 14 \cite{grace2008experiences}. It is likely that adding extra nodes may improve the system's safety due to the presence of backup nodes and providing wider network coverage. This in turn promotes the accuracy of flood prediction, satisfying our main business driver, thus sustaining the GridStix software. Figure~\ref{fig:utility_latency} envisages the utility gained versus reporting latency in accordance to offering up to 20 nodes. This graph is based on the maximum utility achieved from applying a single decision from the set of prior DADs.



\begin{figure}[!t]
\centering
\includegraphics[scale=0.3]{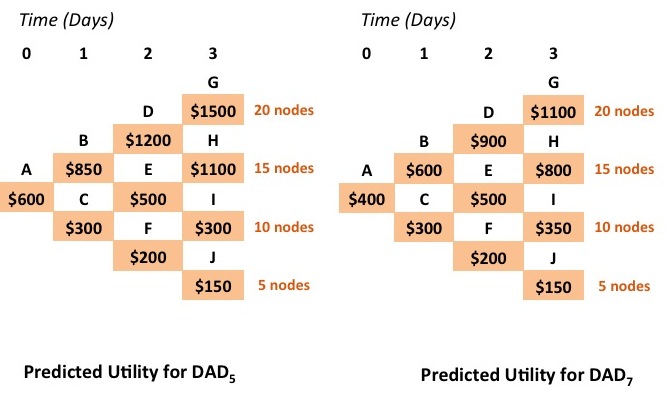}
\caption{Anticipated Values for the utility DAD\textsubscript{5} and DAD\textsubscript{7} due to uncertainty of number of nodes}
\label{fig:bin1}
\end{figure}

\begin{figure}[!t]
\centering
\includegraphics[scale=0.3]{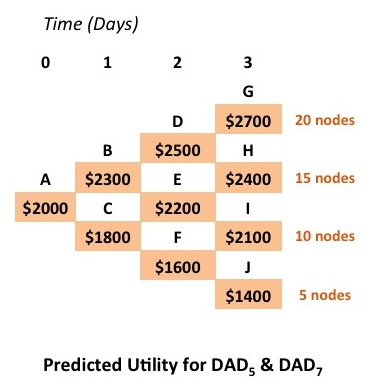}
\caption{Anticipated Values for the utility of implementing both DAD\textsubscript{5} and DAD\textsubscript{7} due to uncertainty of number of nodes}
\label{fig: bin2}
\end{figure}

\paragraph{Favourable Diversification Outcome}
By moving back to case study, DAD\textsubscript{5} and DAD\textsubscript{7} are used for method evaluation. The predicted utility values for the sole implementation of DAD\textsubscript{5} and DAD\textsubscript{7} is revealed in figure~\ref{fig:bin1}. While the deployment of both DADs is shown in figure~\ref{fig: bin2}. Decision makers can vary the base value at cell A (guided by the chart in figure ~\ref{fig:utility_latency}) to perform what-if analysis. A bound of possible values, which can constitute the base value ranging from \$300 to \$2200, where the worst case is \$300 and best case is \$2200. So each DAD will have a different likelihood initial value with respect to its general utility. 

 

\begin{figure}[!t]
\centering
\includegraphics[scale=0.3]{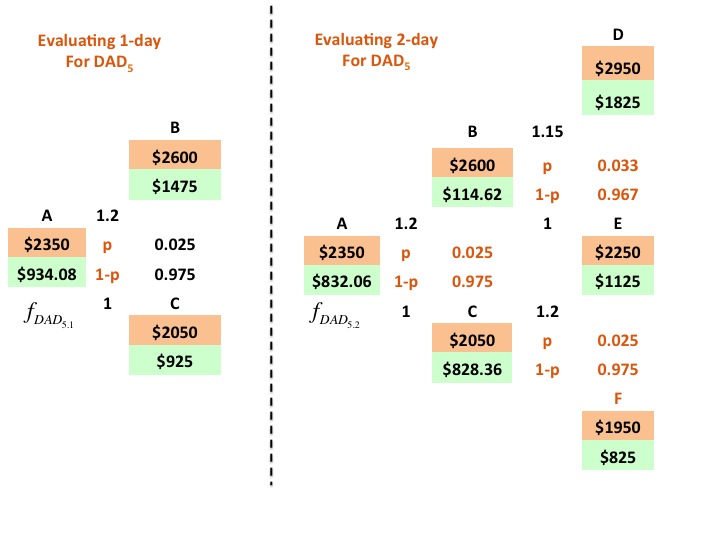}
\caption{Valuation of Options for DAD\textsubscript{5} in case of implementing additional nodes for 1-day and 2-day}
\label{fig: dad_5_1}
\end{figure}

\begin{figure}[!t]
\centering
\includegraphics[scale=0.3]{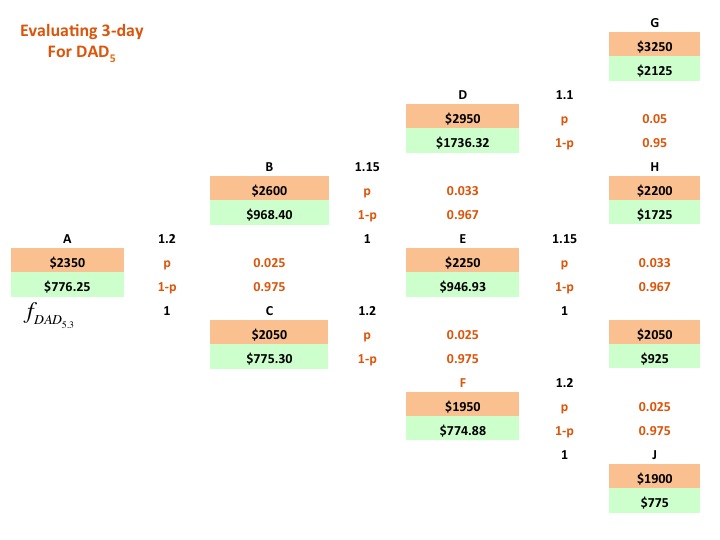}
\caption{Valuation of Options for DAD\textsubscript{5} in case of implementing additional nodes for 3-day}
\label{fig: dad_5_2}
\end{figure}

\begin{figure}[!t]
\centering
\includegraphics[scale=0.3]{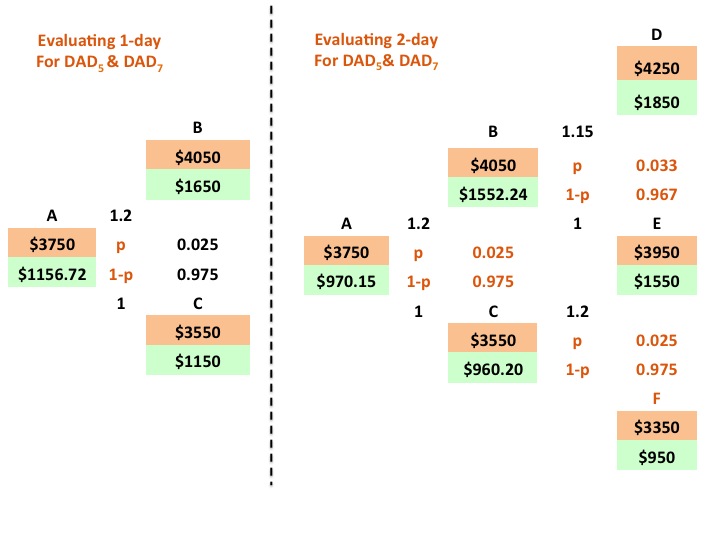}
\caption{Valuation of Options for the deployment of both DAD\textsubscript{5} and DAD\textsubscript{7} for 1-day and 2-day}
\label{fig: dad_5_7_1}
\end{figure}

Moreover, the valuation of options for DAD\textsubscript{5} over varying time slots for uncertainty of implementing additional nodes is clearly shown in figure~(\ref{fig: dad_5_1},\ref{fig: dad_5_2}). Note that the \textit{r} is always 0.5\% in our case and \textit{V\textsubscript{s}} is \$1750.
For detailed analysis, consider cell D for the evaluation of two-day as presented in figure~\ref{fig: dad_5_7_1}, which is the upper cell value:
\begin{equation*}
S_{uu}(2) = V_{s} + V_{DAD_{5}}(2) = 1750 + 1200= 2950
\end{equation*}
The lower cell value is computed as follows:
\begin{equation*}
f_{uu}(2)= max(0, S_{uu}-Cost({DAD_{5}})
\end{equation*}
\begin{equation*}
= max (0, 2950-1125) = 1825 
\end{equation*}
Therefore, the option price formula \textit{f} of DAD\textsubscript{5} is:
\begin{equation*}
f_{DAD_{5}}= f_{DAD_{5.1}} + f_{DAD_{5.2}}+ f_{DAD_{5.3}}
\end{equation*}
\begin{equation*}
= 934.08+832.06+776.25 = \$2542.39 
\end{equation*}

\begin{figure}[!t]
\centering
\includegraphics[scale=0.3]{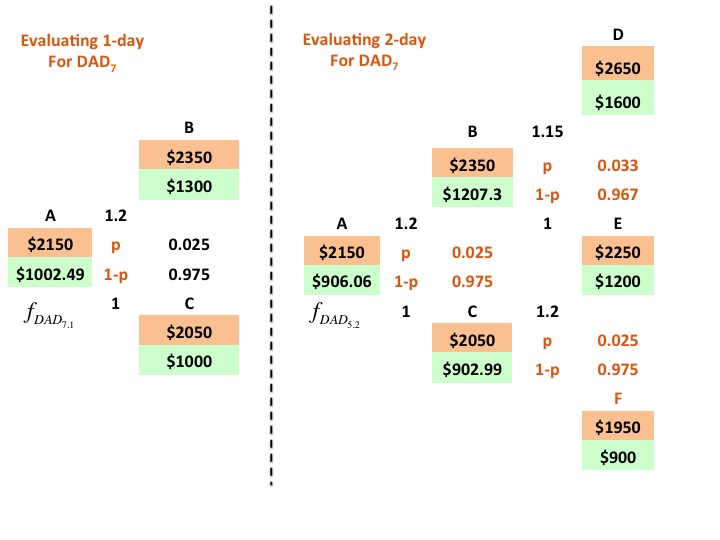}
\caption{Valuation of Options for DAD\textsubscript{7} in case of implementing additional nodes for 1-day and 2-day}
\label{fig: dad_7_1}
\end{figure}

\begin{figure}[!t]
\centering
\includegraphics[scale=0.3]{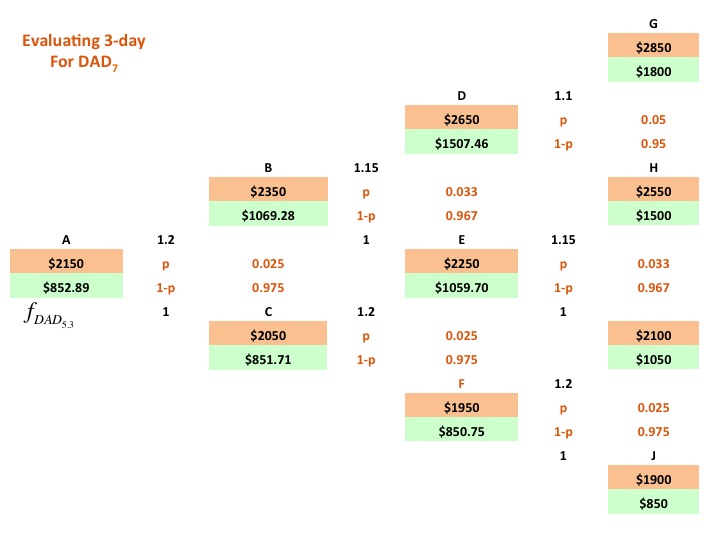}
\caption{Valuation of Options for DAD\textsubscript{7} in case of implementing additional nodes for 3-day}
\label{fig: dad_7_2}
\end{figure}

Same applies for DAD\textsubscript{7}, the valuation of options over varying time slots for uncertainty of implementing additional nodes is shown in figure~(\ref{fig: dad_7_1},\ref{fig: dad_7_2}).
The option price formula \textit{f} of DAD\textsubscript{7} is:
\begin{equation*}
f_{DAD_{7}}= f_{DAD_{7.1}} + f_{DAD_{7.2}}+ f_{DAD_{7.3}}
\end{equation*}
\begin{equation*}
= 1002.49+906.06+852.89 = \$2761.44 
\end{equation*}

\begin{figure}[!t]
\centering
\includegraphics[scale=0.3]{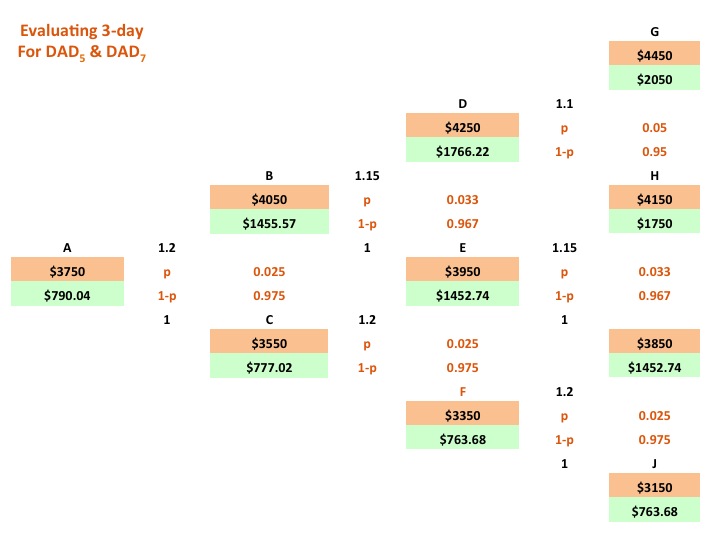}
\caption{Valuation of Options for the deployment of both DAD\textsubscript{5} and DAD\textsubscript{7} for 3-day}
\label{fig: dad_5_7_2}
\end{figure}

By applying diversification, we implemented DAD\textsubscript{5} and DAD\textsubscript{7} together, and then the pay-off is computed to determine whether diversification has yielded favourable or unfavourable outcome. The valuation of options for both is depicted in figure~(\ref{fig: dad_5_7_1},\ref{fig: dad_5_7_2}). The option price value is:

\begin{equation*}
f_{DAD_{5,7}}= 1156.72+970.15+790.04 = \$2916.90 
\end{equation*}

In this context, this clearly shows that the utility has been improved by implementing both DADs simultaneously, which eventually sustains the system. As a result, it is favourable to exercise that option and hence invest in such an architecture. Moreover, the continual monitoring of payoff by waiting was a prominent aid for deciding the most suitable time to switch, as well as an advocate indicator for the system sustainability.
However, this is not the case for all diversification, in the following section, an unfavourable case arose due to diversification.

\begin{figure}[!t]
\centering
\includegraphics[scale=0.3]{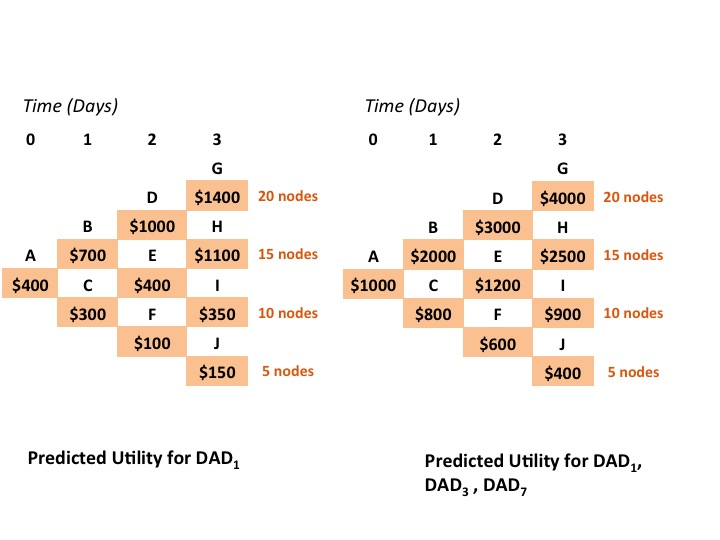}
\caption{Anticipated Values for the utility of implementing DAD\textsubscript{1} vs implementing DAD\textsubscript{1}, DAD\textsubscript{5} and DAD\textsubscript{7} due to uncertainty of number of nodes}
\label{fig:utility_dad1}
\end{figure}

\begin{figure}[!t]
\centering
\includegraphics[scale=0.3]{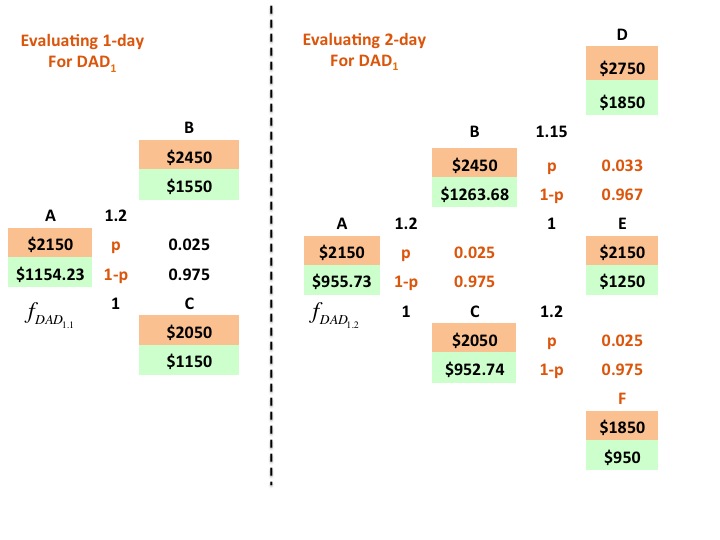}
\caption{Valuation of Options for DAD\textsubscript{1} in case of implementing additional nodes for 1-day and 2-day}
\label{fig: dad_1_1}
\end{figure}

\begin{figure}[!t]
\centering
\includegraphics[scale=0.3]{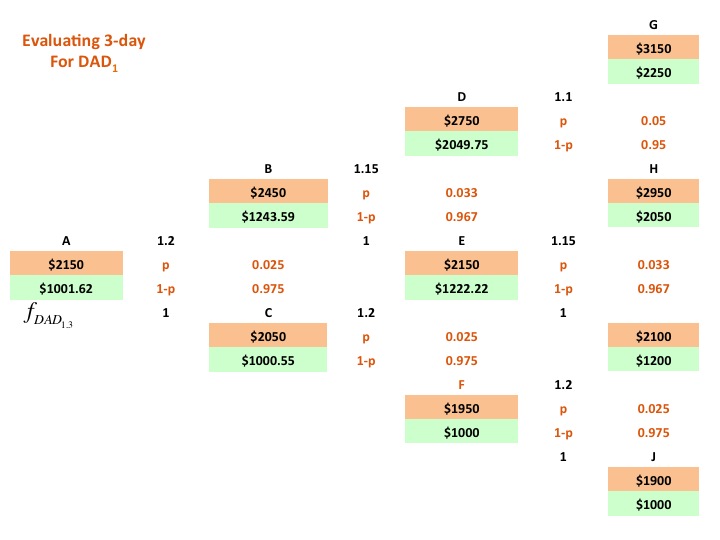}
\caption{Valuation of Options for DAD\textsubscript{1} in case of implementing additional nodes for 3-day}
\label{fig: dad_1_2}
\end{figure}

\begin{figure}[!t]
\centering
\includegraphics[scale=0.3]{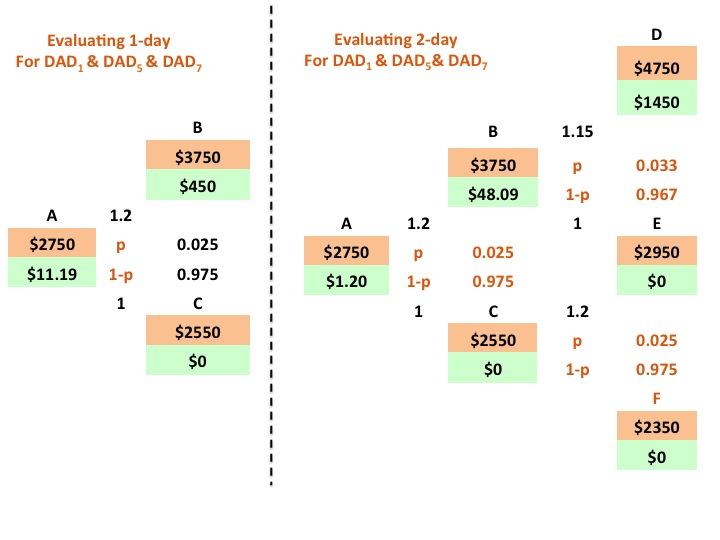}
\caption{Valuation of Options for the deployment of DAD\textsubscript{1}, DAD\textsubscript{5} and DAD\textsubscript{7} for 1-day and 2-day}
\label{fig: dad1_5_7_1}
\end{figure}

\begin{figure}[!t]
\centering
\includegraphics[scale=0.3]{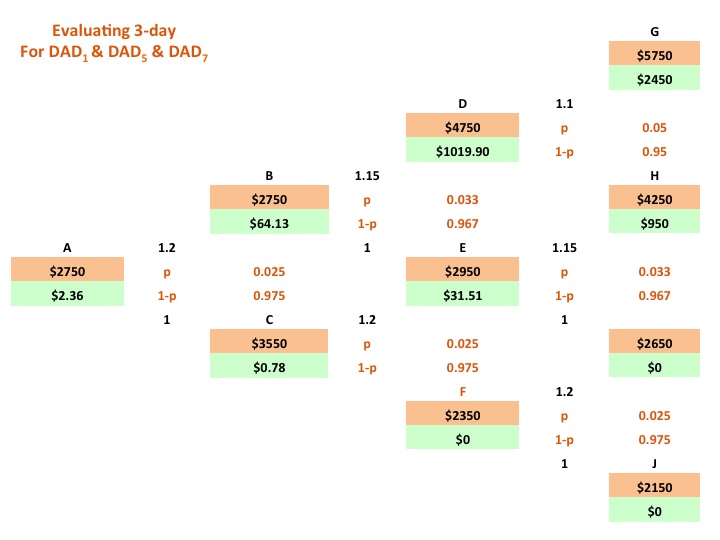}
\caption{Valuation of Options for the deployment of DAD\textsubscript{1}, DAD\textsubscript{5} and DAD\textsubscript{7} for 3-day}
\label{fig: dad1_5_7_2}
\end{figure}

\paragraph{Unfavourable Diversification Outcome}
Typically,we have implemented three DADs to achieve our scenario: DAD\textsubscript{1}, DAD\textsubscript{5}, and DAD\textsubscript{7}. The anticipated utility values for the implementation of DAD\textsubscript{1} versus the three DADs are depicted in figure~\ref{fig:utility_dad1}, as well as the valuation of options for DAD\textsubscript{1} is shown in figure~(\ref{fig: dad_1_1},\ref{fig: dad_1_2}) and for all DADs in figure~(\ref{fig: dad1_5_7_1},\ref{fig: dad1_5_7_2}). By applying the same logic used to calculate the prior option values, \textit{f\textsubscript{DAD\textsubscript{1}}} is \$3111.58 and the option price for 3 DADs equals:  
\begin{equation*}
f_{DAD_{1,5,7}}= 11.19+1.20+2.36 = \$14.75 
\end{equation*}

An unfavourable outcome occurred in subsequent to applying further diversification by implementing additional architectural decisions, as shown by the major drop in option price to \$14.75 and also throughout the time slots a zero payoff appeared repeatedly. Therefore, it is highly desirable to abandon that option. In this regard, this adverse yield may be due to several issues. First, the high costs resulting from the deployment of the three architectural decisions, which outweigh the benefits. Equally importantly, there is interdependence between DADs, for instance both Wifi and FH capabilities are implemented several times, although in reality they will be deployed only once. 
Consequently, we argue that the correlation between the DADs should be studied, in order to provide better results. 

\paragraph{Reflection of the results on sustainability}
This begs to the question: \textit{How decision makers and architects can reason about sustainability benefiting from the above results?}
In our case, we have three possible options. First, the option to switch similar to switching between DADs, when there is a payoff. Second, the option to wait, so that the decision maker analytically monitors the option tree to determine whether it is beneficial to continue or wait until reaching a more profitable outcome. Ultimately, the option to abandon, if the option price kept on falling or having a zero value throughout the time periods, hence it is wiser to    
stop using that DAD and switch to another favourable one. Though we have followed a value-based reasoning to sustainability. The application has focused on the economic performance diversification over time and its contribution to sustainability. On the other hand, the application is versatile as business goal can be still linked to other dimensions of sustainability as described in the manifesto \cite{becker2014karlskrona} and value-based reasoning can be employed as the mechanism to monitor for sustainability.

\section{Related Work}

Over the past decades, in literature, design diversity was mainly envisaged from the context of the design and code levels. These were introduced to achieve some quality goals, such as dependability \cite{bloomfield2003multi}, fault-tolerance \cite{Avizienis:1984:FTD:1319725.1320045}, and security \cite{o2004achieving}. There are two popular methods, N-Version Programming \cite{lyu1992software} and Recovery Block Structure\cite{ammann1988data}, proposed to obtain fault-tolerance.  
Ipek et. al,\cite{ipek2007core} introduced core fusion which applies the software diversity by efficiently providing fine-grain and coarse-grain parallelism, sequential code, and diverse multiprogramming phases. They aim to support dynamic adaptation to requirements via workloads.  
Laperdix et. al, proposed reconfiguration diversified approaches to construct a defence against browser fingerprint tracking \cite{laperdrixmitigating}.  
Obviously, the only closely related work is of Song et. al \cite{songarchitectural}, who recently proposed a diversification architecture ``Diversify" to examine the management of adaptation cost by deploying biological diversification methodologies. Although the model showed promising results in lowering adaptation cost, yet it ignored the interlink to sustainability and value-based perspective. To this extent, none of the approaches addressed it from sustainability point of view. 



Real options approach has been discussed thoroughly in literature in various domains, including software design \cite{sullivan1999software}, software architecture \cite{ozkaya2007quality}, systems design and engineering \cite{baldwin2000design}, software refactoring \cite{sullivan1999software}\cite{bahsoon2005using}, and
COTS-centric development \cite{erdogmus1999quantitative}. Baldwin et. al, \cite{baldwin2000design} enacted the use of real options in systems design and engineering. They empathized on the contribution of modularity on system designs in the form of real options. Similarly, Sullivan et al.\cite{sullivan2001structure}, established the use of real options in software engineering. They proposed an options-based analysis approach to evaluate the use of spiral model in software development \cite{sullivan1999software}\cite{sullivan1996software}. A common goal of model is minimising the uncertainty, providing alternatives, and hence making decision based on prior knowledge from phases. This is analogous to option to defer a decision of investment until a specific optimal time. Ozkaya et. al\cite{ozkaya2007quality}, introduced a model that establish the use of real options in software architecture. In their approach, a variety of different architectural patterns are valued by real options with respect to their relative impact on system quality goals. In other words, the main aim is to determine the most suitable architecture for deployment, given uncertainty future values. Furthermore, Erdogmus et al. \cite{erdogmus1999quantitative} investigated the means of valuing the strategic flexibility in software development using real options.  The main objective is to study the economic incentive of choosing a favourable COTS centric strategy in a project. According to their results, real options theory is preferred over Net Present Value (NPV) analysis, as latter ignores the value of the flexibility in COTS-centric projects. Real options theory was also employed in CBAM aiming to urge about the value of postponing an investment decision in an AS, which in turn enhances CBAM \cite{kazman2001quantifying}. 

\section{Conclusions}



We have described an approach, which makes novel extension of CBAM. The method reasons about diversification in software architecture design decisions using real options. The fundamental premise is that diversification embeds flexibility in an architecture. This flexibility can have value under uncertainty and can be reasoned using Real Options. Real options theory provides an analysis paradigm that emphasizes the value-generating power of flexibility under
uncertainty. The method can be used by the architect and the decision maker to apprise the value of architecting for sustainability via diversification. For instance, the method can be used to inform whether an architecture decision needs to be diversified and to what extent it can benefit from diversification. It also discusses how and when diversification can contribute to long-term value (and when it may cease to add value). Consequently, this value can be used to reflect on sustainability. The application shows that the method can provide systematic assessment for the interlink between sustainability and diversity using value-based reasoning. In a nutshell, architects and decision makers can benefit from the visualisation of the value to assess the extent to which diversification of the architecture design decisions can sustain the software system.

%
\bibliographystyle{abbrv}
\bibliography{phd_ref}  

\end{document}